\documentclass[fleqn,twoside]{article}
\usepackage{espcrc2}

\usepackage{graphicx,epsfig}
\usepackage[figuresright]{rotating}

\newcommand{\AmS}{{\protect\the\textfont2
  A\kern-.1667em\lower.5ex\hbox{M}\kern-.125emS}}

\title{Transport in dimerized and frustrated spin systems}

\author{F. Heidrich-Meisner\address[tubs]{Institut f\"ur Theoretische Physik, Technische Universit\"at Braunschweig,
         \\ Mendelssohnstrasse 3, 38106 Braunschweig, Germany}%
	 \thanks{Corresponding author. Tel.: +49-531-391-5184.
	{\it E-mail:} f.heidrich-meisner@tu-bs.de.},
        A. Honecker\addressmark,
        D.C. Cabra\address[ens]{Laboratoire de Physique, Ecole Normale Superieure de Lyon,\\ 46, All\'{e}e d'Italie, 69364 Lyon, France}
        and
        W. Brenig\addressmark[tubs]}
       
\begin{document}

\begin{abstract}
We analyze the Drude weight for both spin and thermal transport of one-dimensional spin-$1/2$ systems by means
of exact diagonalization at finite temperatures. While the Drude weights are non-zero for finite systems, we find indications of a
vanishing of the Drude weights in the thermodynamic limit for non-integrable models implying normal transport behavior.

\vspace{1pc}
\end{abstract}

\maketitle

\begin{figure}[t]
\texttt{\epsfig{figure=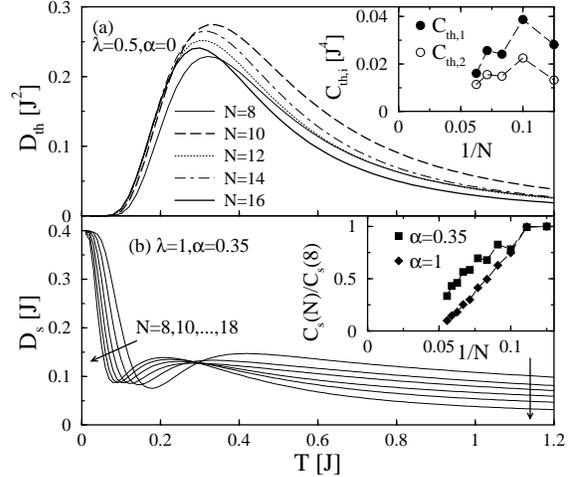,width=7.3cm}}
\vskip-0.6cm
\caption{
(a) Main panel: Thermal Drude weight $D_{\mathrm{th}}(T)$ of a dimerized  chain with $\lambda=0.5$ (see Eq.\ (\ref{eq:1})). Inset: 
 $C_{\mathrm{th},i},i=1,2$ vs. $1/N$ (see text).
(b) Main panel: Drude weight $D_{\mathrm{s}}(T)$ of a frustrated  chain with $\alpha=0.35$. Arrows indicate increasing $N$. 
Inset:  $C_{\mathrm{s}}$ vs. $1/N$ for $N=8,9,\dots,18$.}
\label{fig:1}
\end{figure}

Transport properties of one-dimensional spin-$1/2$ systems have recently attracted
strong interest from the theoretical side (see e.g.\ \cite{kluemper02,gros02,hm02,saito03,orignac02,fujimoto03} and references therein).  An intensively studied issue is the question
under which conditions ballistic transport occurs at zero frequency and finite temperature $T>0$, characterized by a non-zero
Drude weight $D$. This quantity is the zero-frequency weight of the real part of the conductivity $\sigma(\omega)$, 
namely $\mathrm{Re}\, \sigma(\omega)= D(T) \delta(\omega) +\mbox{regular part}$.
While it is known that the anisotropic Heisenberg chain has a finite Drude weight for thermal transport
since the energy-current
operator is a conserved quantity\cite{zotos97}, there is a controversial discussion\cite{gros02,hm02,saito03,orignac02} 
whether transport remains ballistic
if the model is extended by dimerization or frustration.\\
\indent 
In this paper, we will focus on two examples, namely  thermal transport of dimerized chains and 
spin transport in frustrated chains. We consider the  Hamiltonian ($S=1/2$)
\begin{equation}
   H=  \sum_{l=1}^N h_l =  J \sum_{l=1}^N \lbrack \lambda_l\,\vec{S}_l\cdot\vec{S}_{l+1}
              +\alpha\, \vec{S}_l\cdot\vec{S}_{l+2} \rbrack.
\label{eq:1}
\end{equation}
Here, $N$ is the number of sites and $h_l$ is the local energy density.
We set $\lambda_l=1$ if $l$ is even  and $\lambda_l= \lambda$ otherwise. Note that we use periodic boundary conditions.\\
\indent
The thermal Drude weight $D_{\mathrm{th}}(T)$
can be obtained from\cite{zotos97} ($p_n=e^{- E_n/T}/Z; Z=\sum_n e^{- E_n/T}$)
\begin{equation}
D_{\mathrm{th}}(T) =
   \frac{\pi}{N\, T^2} \sum_{m, n\atop E_m=E_n}p_n\,
      |\langle m | j_{\mathrm{th}}|n\rangle|^2 \label{eq:2}
\end{equation}
while the Drude weight $D_{\mathrm{s}}(T)$ for spin transport follows from\cite{shastry90,castella95}
 \begin{equation}
D_{\mathrm{s}}(T) = \frac{\pi}{N}\left\lbrack   \langle - \hat T\rangle - 2\hspace{-0.2cm}\sum_{m, n\atop E_m\not=E_n}\hspace{-0.1cm}p_n\,\frac{|\langle m |
j_{\mathrm{s}}|n\rangle|^2}{E_m-E_n} \right\rbrack.
\label{eq:3}
\end{equation}
$j_{\mathrm{th}}$ and $j_{\mathrm{s}}$ denote the energy- and 
spin-current operator, respectively. They obey 
equations of continuity:
$i \big\lbrack H, h_l[S_l^z]  \big\rbrack= - (j_{\mathrm{th[s]},l+1}-j_{\mathrm{th[s]},l})$ with $j_{\mathrm{th[s]}}=\sum_l j_{\mathrm{th[s]},l}$. 
The local energy density $h_l$ is defined  in Eq.~(\ref{eq:1}) and $S_l^z$ is the local magnetization density.
We refer the reader to  Refs.\ \cite{hm02,bonca94} for  full  expressions of $j_{\mathrm{th[s]}}$.
Both current operators do not commute with $H$ as soon as $\lambda\not= 1$ or $\alpha\not= 0$.
The operator
$ \hat T= J\sum_l\lbrack \lambda_l \,S_l^+S_{l+1}^-+ 4\alpha \,S_l^+S_{l+2}^-+\mbox{H.c.} \rbrack$
 is the kinetic energy\cite{shastry90,bonca94}. \\
\indent 
Now we turn to the discussion of our numerical results for the thermal Drude weight $D_{\mathrm{th}}(T)$ of dimerized chains
which we obtain by complete diagonalization of $H$.  
The data are shown in the main panel Fig.\ \ref{fig:1}(a) for $N=8,\dots,16$ sites. Since finite-size effects are
strongest at low temperatures, we concentrate on $T> J$. Here, the leading contributions to $D_{\mathrm{th}}(T)$
take the form 
\begin{equation}
D_{\mathrm{th}}(T)= C_{\mathrm{th,1}}/T^2+C_{\mathrm{th,2}}/T^3+\dots\,.
\label{eq:5}
\end{equation}
 While we compute $C_{\mathrm{th,1}}$ 
directly by setting $Z=2^N$ and $e^{-\beta E_n}=1$ in Eq.\ (\ref{eq:2}),  $C_{\mathrm{th,2}}$
is extracted by a fit of Eq.\ (\ref{eq:5}) to the numerical data at high $T$. 
As can be seen in the inset of Fig.\ \ref{fig:1}(a), both coefficients exhibit a significant 
decrease  with system size though not monotonically. 
The system sizes 
are too small to draw conclusions about the actual finite-size dependence of $D_{\mathrm{th}}$,
but the observed  behavior does 
not support the notion of  a finite Drude weight for $N\to \infty$  contrasting a recent claim of Ref.\ \cite{gros02}. 
\\ 
\indent
The second example is spin transport of  frustrated chains in the gapped regime.
In the main panel of Fig.\ \ref{fig:1}(b), we show $D_{\mathrm{s}}(T)$ for $\alpha=0.35$ and even $N\leq 18$. The main features 
are  a finite Drude weight at $T=0$ and  a monotonic decrease of $D_{\mathrm{s}}(T)$ for $T>0.5J$. 
$D_{\mathrm{s}}(T=0)>0$ has also been reported in Ref.\ \cite{bonca94} for $\alpha<0.5$ and $N\leq 20$. Extension to 
larger systems at $T=0$ (e.g. with   Lanczos techniques) should clarify whether $D_{\mathrm{s}}(T=0)>0$ 
is a property that survives in the thermodynamic limit $N\to\infty$.\\
\indent
 We further concentrate on the analysis at large $T$, namely the high-temperature prefactor $C_{\mathrm{s}}=\lim_{T\to\infty}
\lbrack T \cdot D_{\mathrm{s}}(T) \rbrack$. Its finite-size dependence is illustrated in the inset of Fig.\ \ref{fig:1}(b) for
$\alpha=0.35,1$ and $N=8,9,\dots,18$. Apart from odd-even finite-size effects, which are more pronounced for $\alpha=0.35$, 
both data sets exhibit a monotonic decrease of $C_{\mathrm{s}}$ with system size, as it is particularly obvious in the case of $\alpha=1$.
This provides  strong evidence for $D_{\mathrm{s}}\to 0$ for $N\to \infty$ in the gapped regime of frustrated chains 
which is consistent with  Ref.\ \cite{rosch00}.\\
\indent
In summary, we have numerically studied the  thermal Drude weight of dimerized chains and the spin Drude weight of 
frustrated chains at $T>0$. The finite-size analysis supports the conclusion of a zero Drude weight  in the thermodynamic 
limit in these examples. 
We have found analogous results for both kinds of transport for the models discussed here and spin ladders which will be reported in
detail elsewhere\cite{hm03a}.

$\mbox{ }$\\


\vspace{-1.5cm}
\begin{thebibliography}{9}
\bibitem{kluemper02} A.~Kl\"umper and K.~Sakai, J. Phys.\ A {\bf 35} (2002)  2173.
\bibitem{gros02} J.~V.~Alvarez and C.~Gros, Phys.\ Rev.\ Lett.\ {\bf 89} (2002) 156603.
\bibitem{hm02} F.~Heidrich-Meisner et al., Phys.\ Rev.\ B {\bf 66} (2002) 140406(R);
  Phys. Rev. Lett {\bf 92}  (2004) 069703.
\bibitem{saito03} K.\ Saito,  Phys.\ Rev.\ B {\bf 67}, (2003) 064410.
\bibitem{orignac02} E.~Orignac et al., Phys.\ Rev.\ B {\bf 67} (2003) 134426. 
\bibitem{fujimoto03} S.~Fujimoto and N.~Kawakami, Phys.\ Rev.\ Lett. {\bf 90} (2003) 197202.
\bibitem{zotos97} X.~Zotos et al., Phys.\ Rev.\ B {\bf 55} (1997) 11029.
\bibitem{shastry90} B.S. Shastry and B. Sutherland, Phys.\ Rev.\ Lett.\. {\bf 65} (1990) 243.
\bibitem{castella95} H. Castella et al., Phys.\ Rev.\ Lett.\. {\bf 74} (1995) 972.
\bibitem{bonca94} J.\ Bon{\v{c}}a et al., Phys.\ Rev.\ B {\bf 50} (1994) 3415.
\bibitem{rosch00} A.\ Rosch and N.\ Andrei, Phys. Rev. Lett. {\bf 85} (2000) 1092.
\bibitem{hm03a} F. Heidrich-Meisner et al., Phys. Rev. B {\bf 68} (2003) 134436.


\end{thebibliography}
\end{document}